\documentclass[modern]{aastex}
\def\bds{M_\textrm{B}}

\def\nk{n_{\rm b}}

\def\csf{\cos f}
\def\cu{\cos u}
\def\su{\sin u}

\def\dert#1#2{\frac{{{\textrm{d}}}{#1}}{{{\textrm{d}}}{#2}}}

\def\eqi{\begin{equation}}
\def\eqf{\end{equation}}
\def\eqia{\begin{eqnarray}}
\def\eqfa{\end{eqnarray}}

\def\rp#1#2{{#1\over#2}}
\def\lb#1{\label{#1}}

\def\bds#1{\boldsymbol{#1}}

\def\co{\cos\omega}
\def\so{\sin\omega}

\def\ton#1{\left(#1\right)}
\def\qua#1{\left[#1\right]}

\def\beq{\begin{equation}}
\def\eeq{\end{equation}}

\def\hb#1{\hat{\mathbf{#1}}}

\def\sb{\overline{s}}

\usepackage{hyperref}
\hypersetup{
    colorlinks=true,
    linkcolor=red,
    citecolor=magenta,
    filecolor=magenta,      
    urlcolor=blue,
    pdftitle={SMV},
    pdfpagemode=FullScreen,}

\usepackage{float,bm}
\usepackage{amsmath,textgreek,w-greek,wasysym}
\usepackage{amssymb,dsfont}
\usepackage{graphicx,epsfig}
\usepackage{txfonts}
\RequirePackage{color}

\allowdisplaybreaks[1]

\begin{document}

\title{Impact of Lorentz violation models on exoplanets dynamics
\vspace{1em}}

\shortauthors{A. Gallerati, M.L. Ruggiero, L. Iorio}

\author{Antonio Gallerati}
\affil{Politecnico di Torino, Dipartimento DISAT, corso Duca degli Abruzzi 24, 10129 Torino, Italy}
\affil{\href{mailto:antonio.gallerati@polito.it}{\texttt{antonio.gallerati@polito.it}}}

\author{Matteo Luca Ruggiero}
\affil{\makebox[1em][c]{Dipartimento di Matematica ``G.Peano'', Universit\`a di Torino, Via Carlo Alberto 10, 10123 Torino, Italy.} \\ INFN - LNL , Viale dell'Universit\`a 2, 35020 Legnaro (PD), Italy}
\affil{\href{mailto:matteoluca.ruggiero@unito.it}{\texttt{matteoluca.ruggiero@unito.it}}}

\author{Lorenzo Iorio}
\affil{Ministero dell'Istruzione, dell'Universit\`{a} e della Ricerca (M.I.U.R.)  \\
Permanent address for correspondence: Viale Unit\`{a} di Italia 68, 70125, Bari (BA)}
\affil{\href{mailto:lorenzo.iorio@libero.it}{\texttt{lorenzo.iorio@libero.it}}}

%

\begin{abstract}
Many exoplanets were detected thanks to the radial velocity method, according to which the motion of a binary system around its center of mass can produce a periodical variation of the Doppler effect of the light emitted by the host star. These variations are influenced by both Newtonian and non-Newtonian perturbations to the dominant inverse-square acceleration; accordingly, exoplanetary systems lend themselves to test theories of gravity alternative to General Relativity. In this paper, we consider the impact of Standard Model Extension (a model that can be used to test all possible Lorentz violations) on the perturbation of radial velocity, and suggest that suitable exoplanets configurations and improvements in detection techniques may contribute to obtain new constraints on the model parameters.
\end{abstract}

\section{Introduction}\lb{sec:intro}

After the first detection of a planet orbiting a main sequence star \citep{Mayor:1995eh}, thousands of exoplanets were detected using different  techninques, such as  radial velocity, transit photometry and timing, pulsar timing, microlensing and astrometry: indeed, each of these techniques is sensitive to specific properties of the planetary systems, unavoidably leading to selection effects in the detection process \citep{perryman,book018}.

Actually, radial velocity (RV) is a powerful tool that is used not only in the search of exoplanets but, more generally, to discover  an invisible celestial object gravitationally bound to another one. The underlying idea is the following: by accurately observing the light spectrum of the visible body, it is possible to detect periodical variations of the wavelength due to Doppler effect  determined by the motion of the system around the center of mass. This is the projection of the  velocity vector onto the line of sight. Obviously, in an exoplanetary system the visible body is the host star, while the invisible one is the planet, but a similar approach can be applied also to a binary system made of a main sequence star and white dwarf, a neutron star or a black hole. 

In a previous work, \citep{Iorio:2017med}, one of the authors introduced a comprehensive approach to obtain the impact of post-Keplerian (pK) corrections to the dominant Newtonian inverse-square acceleration on radial velocity: in particular, they can be of both Newtonian and non-Newtonian origin deriving, for instance, from models of gravity alternative to General Relativity (GR). In fact, on the one hand we know that, more than 100 years after its publication, GR remains the best model to describe gravitation interaction, as its predictions were verified with great accuracy \citep{2014arXiv1409.7871W}. However,  there are challenges coming from the observation of the Universe at very large scales \citet{2016Univ....2...23D} and, in addition, there are known problems when one tries to reconcile GR with the Standard Model  of particle physics. Consequently, it is expected that GR could represent a suitable limit of a more general theory, which we still ignore. As a consequence, there are various and sound motivations to try to extend Einstein's theory: summaries of diverse modified gravity models can be found, e.g., in the review papers \cite{odintsov2007,lobo2008,Tsujikawa:2010zza,harko2011,Capozziello:2011et,CLIFTON2012,Capozziello2015,Berti:2015itd,Cai2016,Nojiri2017,bahamonde21}.

In this context, the role of Lorentz symmetry is quite relevant: in fact, it represents a fundamental property of the mathematical model of spacetime at the basis of GR. The search for a more fundamental theory, hence, brings about a careful investigation of possible Lorentz violations (LV).  The Standard Model Extension (SME) is a framework that can be used to experimentally test all possible deviations Lorentz violations \citep{RevModPhys.83.11,PhysRevD.83.016013,PhysRevD.74.045001,PhysRevD.71.065008,PhysRevD.69.105009,PhysRevD.66.056005,PhysRevD.58.116002,PhysRevD.55.6760,PhysRevD.40.1886,PhysRevD.39.683}.

The purpose of this paper is to calculate the perturbation of radial velocities within the SME, in order to evaluate their impact on current observation of exoplanets.
More specifically, we wish to explore the potential that such a method may have in  constraining relevant LV-parameters in light of the current and expected accuracies in measuring exoplanetary RVs

The paper is organised as follows: we define the impact of the SME coefficients on the system dynamics in Section \ref{sec:def}, and we calculate, to lowest order in eccentricity, the perturbation of the radial velocity in Section \ref{sec:radial}; discussion and conclusion are eventually in Section \ref{sec:disconc}.



%
%
%
%

\section{Definition of the perturbing acceleration} \label{sec:def}

The SME is based on Riemann-Cartan spacetime; in particular (see e.g. \citet{bailey2006signals})   if we focus on the pure-gravity sector, the relevant equation of motion can be derived from a Lagrangian in the form $ L =  L_{\rm LI} + L_{\rm LV}$, where $ L_{\rm LI},  L_{\rm LV}$ refer to the Lorentz-inviariant and Lorentz-violating terms respectively. In the limit of Riemannian spacetime, the pure-gravity sector Lagrangian turns out to be the usual Einstein-Hilbert action $L_{\rm LI} = \sqrt{-g} (R-2\Lambda)/16\pi G$, where $G$ is the Newtonian constant of gravitation, $R$ is the Ricci scalar, $g$ is the metric determinant and $\Lambda$ is the cosmological constant. Then, the Lorentz-violating Lagrangian turns out to be \citep{PhysRevD.69.105009,PhysRevD.74.045001}:

\begin{equation}
  L_{\rm LV} = \frac{\sqrt{-g}}{16\pi G} \left( - u\,R +
  s^{\mu\nu}\,R_{\mu\nu}^{\rm T} + t^{\kappa\lambda\mu\nu}
  C_{\kappa\lambda\mu\nu} \right). \label{eq:LV}
\end{equation}

In the above expression, $R^{\rm T}_{\mu\nu}$ is the trace-free Ricci tensor and $C_{\kappa\lambda\mu\nu}$ is the Weyl conformal tensor. The $u$, $s^{\mu\nu}$ and $t^{\kappa\lambda\mu\nu}$ objects are Lorentz-violating fields: more precisely, they violate the particle local Lorentz invariance and the diffeomorphism, while the observer local Lorentz invariance is maintained \citep{PhysRevD.69.105009}. The post-Newtonian analysis of the SME equations for the pure-gravity sector, as discussed by \citet{PhysRevD.74.045001}, show that the relevant terms in the metric that describe the leading observable effects are determined by the  components of a trace-free matrix $\bar s^{\mu\nu}$,
which are the (rescaled) vacuum expectation values of $s^{\mu\nu}$. It is relevant to point out that, while $\bar s^{\mu\nu}$ is observer Lorentz invariant, it turns out to be particle Lorentz-violating \citep{PhysRevD.69.105009}. Accordingly, it is important to specify the observer reference system that we are using. To begin with, we refer to reference frame at rest with the binary system barycenter. In this frame, according to \citet{bailey2009time,bailey2010lorentz}, it is possible to write the acceleration acting on a test particle in terms of the gravitoelectric field $\bds{a}_{GE}=\bds E_{G}$, where
\beq
E_G^j = -\frac{GM}{r^2}\, \left[\,\hat r^j \left(1+\frac 32\, \sb^{00}
+ \frac 32\, \sb^{kl}\, \hat r^k\, \hat r^l\right) -\sb^{jk}\, \hat r^k \,\right]. \label{eq:defE_{G}}
\eeq
In the above expression, $M$ is the mass of the primary,  $\bds r$ is the position vector with respect to the primary and $\hb r$ is the unit vector, $\displaystyle \hb r=\frac{\bds r}{r}$. Eq. (\ref{eq:defE_{G}}) can be written in the form
\beq
\bds E_{G}=\bds E_{G,N}+\Delta \bds E   , \label{eq:defacc}
\eeq
where $\bds E_{G,N}$ is the Newtonian field, and $\Delta \bds E$ is the perturbation, that can be written as 
\beq
\Delta E^{i}=-\frac{GM}{r^{2}}\,\left(\frac 32\, \sb^{00}+\frac 3 2\, \sb^{kl}\,\hat r^{k}\,\hat r^{l}  \right)\,\hat r^{i}+\frac{GM}{r^{2}}\,\sb^{ik}\,\hat r^{k}. \label{eq:DPerturb}
\eeq

\begin{figure}[t]
\begin{center}
\includegraphics[scale=.40]{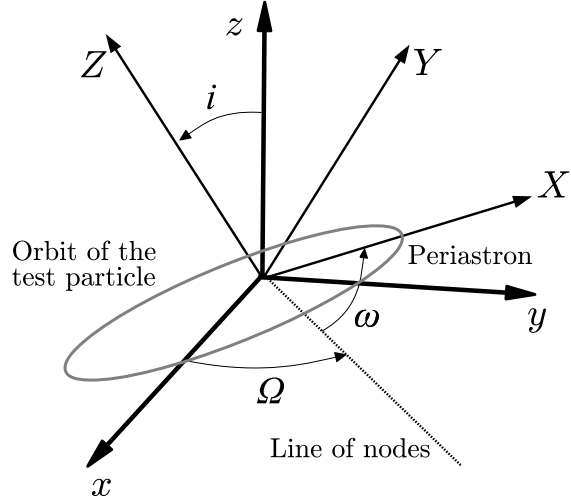}
\caption{Unperturbed orbit of the test particle} \label{fig:orbita}
\end{center}
\end{figure}

In order to evaluate the impact of the above acceleration on the motion of a planet, that can be thought of as a test particle, let us start by suitably parameterizing its unperturbated motion. We first define a reference frame, with origin in the focus of the planet orbit; in this frame, we consider a set of Cartesian coordinates $\{x,y,z\}$, where $z$ is directed along the line of sight toward the Earth. An arbitrary configuration of the test particle orbit is shown in  Figure \ref{fig:orbita}: besides the already mentioned Cartesian coordinate system $\{x,y,z\}$, with unit vectors  $\bds u_{x}, \bds u_{y}, \bds u_{z}$,  we introduce another Cartesian coordinate system $\{X,Y,Z\}$, with the same origin, and unit vectors $\bds u_{X}, \bds u_{Y}, \bds u_{Z}$. The orbital plane is the $XY$ plane, and we denote with $\Omega$  the angle between the $x$ axis and the line of the nodes, while the angle between the $z$ and $Z$ axes is called $i$. The periastron is along the $X$ axis, and we denote by $\omega$ the argument of the periastron, i.e. the angle between the line of nodes and the $X$ axis. 
The following relations hold between the unit vectors of the two Cartesian coordinate systems (see e.g. \citet{bertotti}):
\begin{eqnarray}
\bds u_{X} & = & \left(\cos \omega \cos \Omega-\sin \omega \cos i \sin \Omega \right) \bds u_{x}+\left(\cos \omega \sin \Omega+\sin \omega \cos i \cos \Omega \right) \bds u_{y}+\sin \omega \sin i  \, \bds u_{z} \nonumber \\
\bds u_{Y} & = & \left(-\sin \omega \cos \Omega-\cos \omega \cos i \sin \Omega \right) \bds u_{x}+\left(-\sin \omega \sin \Omega+\cos \omega \cos i \cos \Omega \right) \bds u_{y}+\cos \omega \sin i  \, \bds u_{z} \nonumber \\
\bds u_{Z}&=& \sin i \sin \Omega \, \bds u_{x}-\sin i \cos \Omega \, \bds u_{y}+ \cos i\, \bds u_{z}. \label{eq:trasxyzXYZ}
\end{eqnarray}
In what follows, for direct comparison with previous works (see e.g. \citet{bailey2006signals}), we will use the following notation for the above vectors:
\beq
\bds u_{X} \rightarrow \bds P, \quad \bds u_{Y} \rightarrow \bds Q, \quad \bds u_{Z} \rightarrow \bds N. \label{eq:defPQ}
\eeq
Notice that $\bds P$ is directed from the focus (and origin of the coordinate system) to the periastron; $\bds N$ is orthogonal to the orbital plane. The unit vectors $\bds P, \bds Q, \bds N$ depends on the orbital elements only. Let $\bds X$ denote the position vector of the test particle which, in the orbital plane can be written as 
\beq
\bds X = r(f) \cos f \, \bds P+ r(f) \sin f \, \bds Q, \label{eq:motionR}
\eeq
where the Keplerian ellipse, parameterized by the true anomaly $f$,  is written as
\beq
r(f)=\frac{a\left(1-e^{2}\right)}{1+e\cos f}\, 
\label{eq:keplerorbit1}
\eeq
in terms of the semi-major axis $a$ and eccentricity $e$.

Along the orbit, we define the radial vector $\bds R$ 
\beq
\bds R =   \cos f \, \bds P+  \sin f \, \bds Q, \label{eq:defR}
\eeq
and the transverse vector $\bds T$
\beq
\bds T =   -\sin f \, \bds P+  \cos \, \bds Q. \label{eq:defT}
\eeq

The perturbing acceleration (\ref{eq:DPerturb}) must be evaluated along the orbit, so $\hb r=\bds R$. Accordingly, using the definitions (\ref{eq:defR}), (\ref{eq:defT}), the perturbing acceleration can be written in the form
\beq
\Delta E^{i}=-\frac{GM}{r^{2}}\, \alpha_{1}\,\hat r^{i}+\frac{GM}{r^{2}}\,\sb^{ik}\,\hat r^{k}, \label{eq:pertA2}
\eeq
where
\beq
\alpha_{1}=\frac 3 2 \left[\sb^{00}+\sb^{PP}\cos^{2}f+\sb^{QQ}\sin^{2}f+2\,\sb^{PQ}\sin f\cos f \right],
\eeq
with 
\beq
\sb^{PP}=\sb^{kl}\hat P^{k} \hat P^{l}, \quad \sb^{QQ}=\sb^{kl}\hat Q^{k} \hat Q^{l}, \quad \sb^{PQ}=\sb^{kl}\hat P^{k} \hat Q^{l}. \label{eq:defprog1}
\eeq
Notice that $\sb^{PP}$, $\sb^{QQ}$, $\sb^{PQ}$ depend on the orbital elements only.

Now, we can calculate the radial, transverse and normal components of the perturbing acceleration (\ref{eq:pertA2}).
We obtain
\begin{align}
E_{r}&=\Delta \bds E \cdot \bds R=-\frac{GM}{r^{2}}\,\left[\frac 3 2 \,\sb^{00}+\frac 1 2 \left( \sb^{PP}\cos^{2}f+\sb^{QQ}\sin^{2}f+2\sb^{PQ}\sin f\cos f  \right) \right], \label{eq:defEr}
\\[\jot]
E_{t}&=\Delta \bds E \cdot \bds T=\frac{GM}{r^{2}}\,\left[\cos f \sin f\,\left(\sb^{QQ}-\sb^{PP}\right)+\left(\cos^{2}f-\sin^{2}f \right)\,\sb^{PQ} \right], \label{eq:defEt}
\\[2\jot]
E_{n}&=\Delta \bds E \cdot \bds N=\frac{GM}{r^{2}}\,\left[\cos f\,\sb^{NP}+\sin f\,\sb^{NQ} \right], \label{eq:defEn}
\end{align}
where
\beq
\sb^{NP}=\sb^{kl}\hat N^{k} \hat P^{l}, \quad \sb^{NQ}=\sb^{kl}\hat N^{k} \hat Q^{l}. \label{eq:defprog2}
\eeq

Given the components of the perturbing acceleration, we may write the Gauss equations for the
variations of the semi-major axis $a$, the eccentricity $e$, the inclination $i$, the longitude of the ascending node $\Omega$, the
argument of pericentre $\omega$ and the mean anomaly $\mathcal{M}$:
\begin{align}
\dert{a}{t} & = \rp{2}{{\nk}\sqrt{1-e^2}}  \left[\,e\,E_{r}\,\sin f+E_{t}\left(\rp{p}{r}\right)\,\right],
\lb{smax}
\\[\jot]
\dert{e}{t} & = \rp{\sqrt{1-e^2}}{{\nk}\,a}\left\{\,E_{r}\sin f+E_{t}\left[\cos f+\rp{1}{e}\left(1-\rp{r}{a}\right)\right]\,\right\},
\\[\jot]
\dert{i}{t} & = \rp{1}{{\nk}\,a\sqrt{1-e^2}} \,E_{n}\,\left(\rp{r}{a}\right)\,\cos (\omega+f),
\lb{in}
\\[\jot]
\dert{\Omega}{t} & = \rp{1}{{\nk}\,a\,\sin i\,\sqrt{1-e^2}}\, E_{n}\,\left(\rp{r}{a}\right)\,\sin (\omega+f),
\lb{nod}
\\[\jot]
\dert{\omega}{t} & = -\cos i\,\dert{\Omega}{t}+\rp{\sqrt{1-e^2}}{{\nk}\,a\,e}\left[-E_{r}\cos f+E_{t}\,\left(1+\rp{r}{p}\right)\,\sin f\right],
\lb{perigeo}
\\[\jot]
\dert{\mathcal{M}}{t} & = {\nk} -\rp{2}{{\nk}\,a}\,
E_{r}\,\left(\rp{r}{a}\right)-\frac{1-e^{2}}{\nk\,a\,e}\left[-E_{r}\cos f+E_{t}\,\left(1+\frac r p \right)\,\sin f \right].
\lb{manom} 
\end{align}
In the above equations, ${\nk}=2\pi/T$ is the Keplerian mean motion%
\footnote{For an
unperturbed Keplerian ellipse in the gravitational field of a body with mass $M$,  it is ${\nk}=\sqrt{GM/a^3}$.},
$T$ is the test particle's orbital period, $p=a\left(1-e^{2}\right)$ is the semilatus rectum.

In summary, in order to calculate the variation of orbital elements, we must evaluate the perturbing acceleration  onto the unperturbed Keplerian ellipse, and then it must be inserted into Eqs.(\ref{smax})-(\ref{manom}); then, we must average over one
orbital period $T$. To this end, the following relation  
\beq 
dt = \frac{(1-e^2)^{3/2}}{\nk\,(1+e\cos f)^2}\, df \label{eq:dtdf00} 
\eeq
will be used. 


\section{The method of radial velocity}\label{sec:radial}

As discussed by \citet{Iorio:2017med}, the presence of a perturbing acceleration, whatever its origin is (Newtonian or non-Newtonian), modifies the velocity vector $\mathbf v$ of the motion of the test particle  relative to its primary.%
\footnote{Notice that all the following results hold for the binary's relative orbit; the resulting shift  for the stellar RV can be straightforwardly obtained by rescaling the final formula by the ratio of the planet's mass  to the sum  of the masses of the parent star and of the planet itself.} 
Namely (see also \citet{casotto1993position}), the instantaneous changes  $\Delta\mathrm{v}_{R},~\Delta\mathrm{v}_{T},~\Delta\mathrm{v}_{N}$  of the radial, transverse and out-of-plane components $\mathrm{v}_{R},~\mathrm{v}_{T},~\mathrm{v}_{N}$ are
\begin{align}
\Delta \mathrm{v}_{R}\ton{f} \lb{vR} =\nonumber &-\rp{\nk\,a\,\sin f}{\sqrt{1-e^2}}\qua{\rp{e}{2a}\,\Delta a\ton{f} + \rp{a}{r\ton{f}}\,\Delta e\ton{f} } -\rp{\nk \,a^3}{r^2\ton{f}}\,\Delta{\mathcal{M}}\ton{f}  - 
\\[\jot]
&-\rp{\nk \,a^2}{r\ton{f}}\sqrt{1-e^2}\,\qua{\cos i\, \Delta\Omega\ton{f} + \Delta\omega\ton{f}}, 
\\[3ex]
\Delta \mathrm{v}_{T}\ton{f} \lb{vT} = &-\rp{\nk \,a\sqrt{1-e^2}}{2r\ton{f}}\,\Delta a\ton{f} + \rp{\nk\, a \ton{e+\cos f} }{\ton{1-e^2}^{3/2}}\,\Delta e\ton{f} +\rp{\nk \,a\, e \sin f}{\sqrt{1-e^2}}\qua{\cos i \,\Delta\Omega\ton{f} + \Delta\omega\ton{f}}, 
\\[1ex]
\Delta \mathrm{v}_{N}\ton{f} \lb{vN} = \:&\rp{\nk \,a}{\sqrt{1-e^2}}\qua{\ton{\cos u + e \cos\omega}\Delta i\ton{f} + \ton{\sin u + e \so}\sin i\,\Delta\Omega\ton{f} }.
\end{align}

In the above equation, there are the instantaneous changes of the Keplerian orbital elements $\Delta a\ton{f},~\Delta e\ton{f},~\Delta i\ton{f},~\Delta\Omega\ton{f},~\Delta\omega\ton{f}$; they can be calculated using the general relation
\eqi
\Delta\kappa\ton{f}=\int_{f_0}^f\dert{\kappa}{t} \,\dert{t}{f^{'}} \,df^{'},
\qquad \kappa=a,~e,~i,~\Omega,~\omega,\lb{Dk}
\eqf
where the time derivatives can be obtained from the Gauss equations (\ref{smax})-(\ref{manom}),  $\tfrac{dt}{\,df^{'}}$ from Eq. (\ref{eq:dtdf00}) and $f_{0}$ is the expression of the true anomaly at a given epoch.

Care must paid to the mean anomaly since, as pointed out by \cite{Iorio:2017med}, the possible change of the mean motion $\nk$ can influence the variation of the mean anomaly: in particular, this may happen when the perturbing acceleration provokes a variation of the semimajor axis. As shown by \citet{PhysRevD.74.045001}, this is not the case for the modified gravity models that we are dealing with.

Then, it is possible to obtain the instantaneous change experienced by the radial velocity by taking the $z$ component of the perturbation of the relative velocity  $\displaystyle \Delta \mathbf{v} = \Delta \mathrm{v}_{R}~\bds{R} + \Delta \mathrm{v}_{T}~\bds{T} + \Delta \mathrm{v}_{N}~\bds{N}$. Accordingly, we get
\begin{align}
\Delta \mathrm{v}_z\ton{f}\nonumber =  &-\rp{\nk\sin i \ton{e\co + \cu}}{2\sqrt{1-e^2}}\,\Delta a\ton{f} + 
\\[2\jot]
&+ \rp{\nk\,a\sin i\,\big\{ 4\cos\ton{2f +\omega} + e\qua{-\cos\ton{f-\omega}  +  4\cu +\cos\ton{3f + \omega}  }  \big\}}{4\ton{1-e^2}^{3/2}}\,\Delta e\ton{f} + 
\\[\jot]
&+ \rp{\nk\,a\cos i \ton{e\co+\cu}}{\sqrt{1-e^2}}\,\Delta i\ton{f} - \rp{\nk\,a\sin i \ton{e\so + \su}}{\sqrt{1-e^2}}\,\Delta \omega\ton{f} -
\\[\jot]
&- \rp{\nk\, a\ton{1+e\csf}^2\sin i \su }{\ton{1 - e^2}^2}\,\Delta\mathcal{M}\ton{f}, \lb{Dvzf}
\end{align}
where $u\doteq \omega + f$  is the argument of latitude.

Using Eq. (\ref{Dvzf}), we can calculate the net change of the radial velocity over an  orbital period, namely
\begin{equation}
\left\langle\Delta \mathrm{v}_z\right\rangle = \, \frac{1}{T}\!\int\limits_{f_0}^{f_0+2\pi}\!\frac{dt}{df}\;\Delta v_z\;df\:. \label{eq:deltavzavg}
\end{equation}
The explicit expression of $\left\langle\Delta \mathrm{v}_z\right\rangle$  can be calculated, but it will not be displayed here since it is quite unmanageable; rather, to evaluate its magnitude, we perform an expansion in powers of the eccentricity $e$, and write the lowest order terms. Accordingly, we obtain
\begin{equation}
\left\langle\Delta \mathrm{v}_z\right\rangle \simeq \frac{a}{T}\,\left(c_0+c_1\:e\right),  \label{eq:deltavzavg1}
\end{equation}
where
\begin{equation}
\begin{split}
c_0=\frac{\pi\,}{2}\,\Big\{
&\,2\,\cos i\,\left[s_{NP}\,\sin f_0-s_{NQ}\,\cos f_0\right]+\\
&+\sin i\,\cos(f_0+\omega)\left[\,3\,(s_{PP}-s_{QQ})\,\cos(2f_0)+
2\,\Big(6\,s_{00}+s_{PP}+s_{QQ}+3\,s_{PQ}\,\sin(2f_0)\Big)\,\right]\,
\Big\}\:,
\end{split}
\end{equation}
and
\begin{equation}
\begin{split}
c_1=\frac{\pi\,}{8}\,\Big\{&\,
8\,\cos i\,\cos f_0\,\left[s_{NQ}\,\cos f_0 - s_{NP}\,\sin f_0\right]\:+
\\
&+\sin i\,\Big[\,9\,(s_{PP}-s_{QQ})\,\cos(2f_0- \omega)+14\,(6\,s_{00}+s_{PP}+s_{QQ})\,\cos(\omega)+
\\
&\,\qquad\quad +2\,(18\,s_{00}+7\,s_{PP}-s_{QQ})\,\cos(2f_0+\omega)+3\,(s_{PP}-s_{QQ})\,\cos(4f_0+\omega)+
\\
&\,\qquad\quad +2\,s_{PQ}\,\big(9\,\sin(2f_0-\omega)+8\,\sin(2f_0+\omega)+3\,\sin(4f_0+\omega)\big)
\Big]\:
\Big\}\:.
\end{split}
\label{eq:c1}
\end{equation}
The above expressions suggest there is a non null net change also at zeroth order in the eccentricity. We notice that
the change in the radial velocity can be expressed in the form 
\beq
\left\langle\Delta \mathrm{v}_z\right\rangle \simeq v_{m}\,\Delta S, \label{eq:defpertuno}
\eeq
where $v_{m}$ is the mean orbital speed and $\Delta S$ is a factor which is linear depending on the elements of the Lorentz-violating matrix $\bar s^{\mu\nu}$ and, in addition, it depends on the (bounded) trigonometric function of the angular orbital elements.

\section{Discussion and conclusion}\label{sec:disconc}

As we showed above, the presence of Lorentz-violating terms will produce variation of the radial velocity that can be written (see Eq. \eqref{eq:defpertuno}) in the form: \begin{equation*}
\mathrm{mean\ orbital\ speed}\:\times\:\mathrm{perturbation}\,.
\end{equation*} We emphasise that, even though we refer to exoplanetary systems, these results can be applied to any gravitationally bound binary system.

The elements of the Lorentz-violating matrix $\bar s^{\mu\nu}$ were estimated from different tests, including  atomic gravimetry, Lunar Laser Ranging, Very Long Baseline Interferometry, planetary ephemerides, Gravity Probe B, binary pulsars, high energy cosmic rays (see e.g. \citep{hees_universe} and references therein). In particular, \citet{hees_universe}  (see Table 9) report a combined analysis of the best constraints deriving from various observations and experiments, and  the results range from $10^{-12}$ up to  $10^{-5}$. In this regard, using the above \eqref{eq:deltavzavg1}--\eqref{eq:c1}, for suitable systems featuring small eccentricity, perturbations of the order of  $\,\sim \mathrm{m}\,\mathrm{s}^{-1}$ or bigger can be found.   

At this point, a comment should be done: as we said in Section \ref{sec:def}, in this context it is of utmost importance to specify the reference frame considered and, in our derivation, we referred to the reference frame at rest with the system barycenter. However, the above constraints refer to an asymptotically inertial frame co-moving with the Solar System: as a consequence, as  discussed by  \cite{shao}, to relate the two frames a Lorentz transformation is required which, due to the smallness of the relativity velocity of the planetary system with respect to the Solar System, can be considered as a pure rotation. Accordingly, we do not expect that these coordinate transformations could significantly change the order of magnitude of the estimates of the Lorentz-violating terms. In any case, once that the planetary system is chosen, the transformation can be easily performed to obtain more precise estimates. 

Recent perspectives on radial velocity measurements \citep{fischer2016state,gilbertson20,Matsuo:2021idy}  suggest that a precision of $0.02-0.1\, \mathrm{m\, s^{-1}}$ could be attained in the  near future. Accordingly, if these techniques could be successfully  applied to exoplanetary systems, we would have a new opportunity to explore the impact of SME coefficients outside the Solar System. To this end, with planetary  speeds of the order of $10^{4 } \, \mathrm{m\, s^{-1}}$, constraints of the the order of $10^{-6}$ could be obtained. However, the exploration of exoplanetary systems brings about features that are unexpected on the basis of the knowledge of the Solar System: for instance, there are planets moving at very high speed, of the order of $10^{6 } \, \mathrm{m\, s^{-1}}$ (see \citet{lam2021gj}): a combination of these peculiar planets and improvements in detection techniques could lead to even tighter constraints on the SME coefficients.

\bibliography{SMV}


\end{document}